\documentclass[11pt,twoside]{article}

\usepackage{asp2004}
\usepackage{epsf}
\usepackage{psfig}
\usepackage{lscape}
\usepackage{epsfig}
\begin{document}
\title{Giant Outbursts of the Eta Carinae-P Cygni Type}    %%% Fill in title
\author{Nathan Smith}   %%% Fill in author names
\affil{Center for Astrophysics and Space Astronomy, University
of Colorado, 389 UCB, Boulder, CO 80309, USA;
nathans@astro.berkeley.edu}    %%% Fill in author affiliations

\begin{abstract} %%% Abstract to run on from here.

I discuss the role of short-duration eruptive mass loss in the
evolution of very massive stars.  Giant eruptions of Luminous Blue
Variables (LBVs) like the 19th century event of $\eta$ Carinae can
remove large quantities of mass almost instantaneously, making them
significant in stellar evolution.  They can potentially remove more
mass from the star than line-driven winds, especially if winds are
clumped such that O star mass-loss rates need to be revised downward.
When seen in other galaxies as ``supernova impostors'', these
eruptions typically last for less than a decade, and they can remove
of order 10 M$_{\odot}$ as indicated by massive nebulae around LBVs.
Such extreme mass-loss rates cannot be driven by radiation pressure on
spectral lines.  Instead, these outbursts must either be
continuum-driven super-Eddington winds or outright hydrodynamic
explosions, both of which are insensitive to metallicity.  As such,
this eruptive mode of mass loss could play a pivotal role for massive
metal-poor stars in the early universe.

\end{abstract}
%%% MAIN BODY OF TEXT GOES HERE. CONSULT "INSTRUCTIONS FOR AUTHORS USING
%%% LATEX2E MARKUP", SECTIONS 2.3-2.6 FOR HELP WITH EQUATIONS, FIGURES,
%%% AND TABLES.

\section{Introduction}

The main question I wish to address here is whether the majority of
mass lost during the lifetime of the most massive stars occurs
primarily via steady line-driven stellar winds, or instead through
violent, short-duration eruptions.  The two extremes are shown
graphically in Figure 1.  This is critical for understanding how
$\dot{M}$ scales with metallicity.  In this contribution I draw
attention to the role of LBV eruptions, advocating for their
importance.  The essential points of the argument, already made by
Smith \& Owocki (2006), are the following:

\begin{itemize}

\item Recent studies of hot star winds indicate that MS mass-loss
  rates are lower than previously thought due to the effects of
  clumping (Fullerton et al.\ 2006; Bouret et al.\ 2005; Puls et al.\
  2006; Crowther et al.\ 2002; Hillier et al.\ 2003; Massa et al.\
  2003; Evans et al.\ 2004).  Revised $\dot{M}$ values are inadequate
  to reduce the star's mass enough to reach the WR phase.

\item Observations of nebulae around LBVs and LBV candidates have
  revealed very high ejecta masses -- of order 10 M$_{\odot}$ (Fig.\
  2). Some objects show evidence for multiple shell ejections on
  timescales of 10$^3$ years.  These eruptions could remove a large
  fraction of the total mass of the star.

\item The extreme mass-loss rates of these LBV bursts imply that line
  opacity is too saturated to drive them, so they must instead be
  either continuum-driven super-Eddington winds (see Owocki et al.\
  2004) or outright hydrodynamic explosions.  Unlike steady winds
  driven by lines, the driving in these eruptions may be largely
  independent of metallicity, and might play a role in the mass loss
  of massive metal-poor stars.

\end{itemize}

For reasons that may be obvious, my talk at this meeting and the
original paper (Smith \& Owocki 2006) were deliberately provocative.
Rather than repeat that discussion, I will briefly elaborate on a few
of these issues, and will consider alternatives and further
implications.

\section{Balancing the Budget: LBV Eruptions}

The most likely mechanism to rectify the hefty mass deficit left by
clumped winds is giant eruptions of LBVs (e.g., Davidson 1987; Lamers
1987; Humphreys \& Davidson 1994; Humphreys, Davidson, \& Smith 1999;
Smith \& Owocki 2006), where $\dot{M}$ and L$_{\rm Bol}$ increase
substantially.  While we do not yet fully understand what causes these
giant LBV outbursts, we know empirically that they do indeed occur,
and that they drive substantial mass off the star.  Deduced masses of
LBV and LBV-candidate nebulae are plotted in Figure 2.  For stars with
log(L/L$_{\odot}$)$>$6, nebular masses of 10 M$_{\odot}$ are quite
reasonable, {\it perhaps suggesting that this is a typical mass
ejected in a giant LBV eruption}.

If these large nebular masses are typical for LBV outbursts, then only
a few such eruptions occurring sequentially are needed to remove a
large fraction of the star's mass.  This is shown schematically in
Figure 1 for stars with initial masses of 120 and 60 M$_{\odot}$.
Notice that at 60 M$_{\odot}$, the LBV eruptions are more numerous and
each one is less massive than at 120 M$_{\odot}$; this is mostly
hypothetical, but is based on the presumption that a more luminous
star will have more violent mass ejections because of its closer
proximity to the Eddington limit.  For example, we might expect that
$\eta$ Car currently has an Eddington parameter of $\Gamma$=0.9 or
higher, whereas a less luminous LBV like P Cygni probably has
$\Gamma$=0.5 or so.  It is therefore also likely that the relative
importance of eruptive mass loss diminishes with lower L.  Measuring
the mass ejected in each burst, plus their frequency and total number
as functions of L and Z are probably the most important observations
to unravel the role of LBVs in stellar evolution.
 
Our best example of this phenomenon is the 19th century ``Great
Eruption'' of $\eta$ Carinae.  The event was observed visually, the
mass of the resulting nebula has been measured (12--20 M$_{\odot}$ or
more; Smith et al.\ 2003), and proper motion measurements of the
expanding nebula indicate that it was ejected in the 19th century
event (e.g., Morse et al.\ 2001).  The other example for which this is
true is the 1600 {\sc ad} eruption of P Cygni, although its shell
nebula has a much lower mass (Smith \& Hartigan 2006).  Both
$\eta$~Car and P~Cyg are surrounded by multiple, nested shells
indicating previous outbursts (e.g., Walborn 1976; Meaburn 2001).
While the shell of P~Cyg is less massive than $\eta$~Car's nebula, it
is still evident that P~Cyg shed more mass in such bursts than via its
stellar wind in the time between them (Smith \& Hartigan 2006).

Although LBV eruptions are rare, a number of extragalactic $\eta$ Car
analogs or ``supernova impostors'' have been observed.  Several
massive circumstellar shells have also been inferred to exist around
supernovae and gamma-ray bursters (see Smith \& Owocki 2006 and
references therein).  These indicate that the eruption of $\eta$ Car
is not an isolated, freakish event, but instead may represent a common
rite of passage for the most massive stars. A massive ejection event
may help initiate the LBV phase, by lowering the star's mass, raising
its L/M ratio, and drawing it closer to instability associated with an
opacity-modified Eddington limit (Appenzeller 1986; Davidson 1987;
Lamers \& Fitzpatrick 1988; Humphreys \& Davidson 1994).  In any case,
meager mass-loss rates through stellar winds, followed by huge bursts
of mass loss in violent eruptions at the end of core-H burning (Fig.\
1) may significantly alter stellar evolution models.

% FIGURE 1 ----------
\begin{figure}[!ht]\begin{center}
\epsfig{file=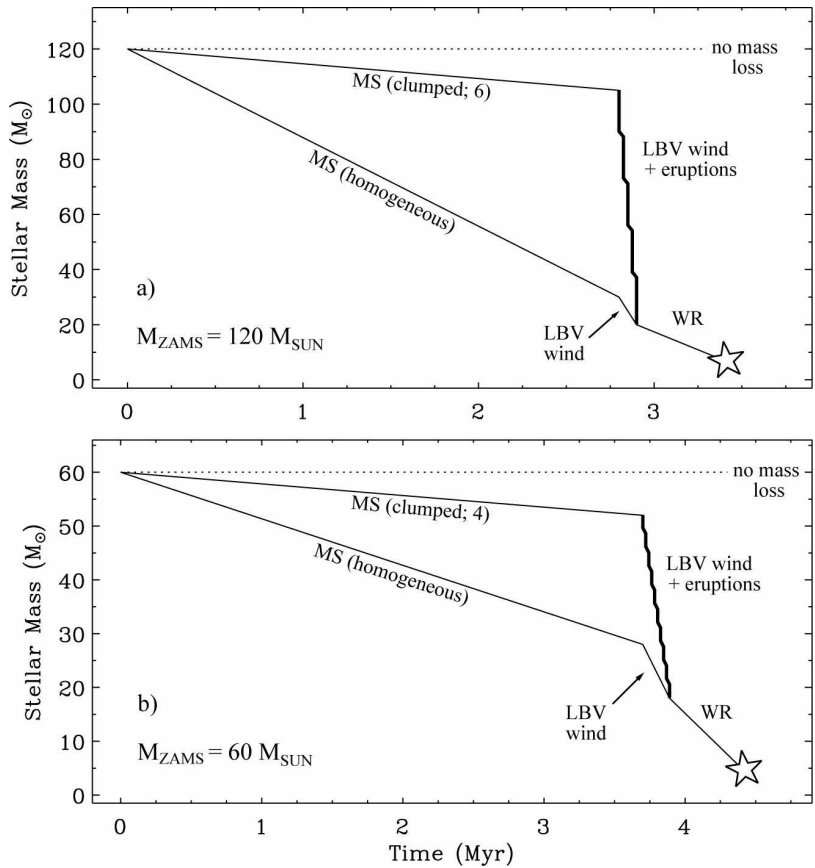,width=4.2in}
%\epsscale{0.99}
%\plotone{lbvneb.eps}
\caption{Schematic representation of a star's mass as a function of
  time.  Two extreme scenarios are shown: One has higher conventional
  O-star mass-loss rates assuming homogeneous winds on the
  main-sequence (MS) with no clumping.  This is followed by a brief
  LBV wind phase and a longer WR wind phase before finally exploding
  as a supernova; this is the type of scenario usually adopted in
  stellar evolution calculations.  The second has much reduced
  $\dot{M}$ on the main sequence (assuming clumping factors of 4--6),
  followed by an LBV phase that includes severe mass loss in brief
  eruptions plus a steady wind; this is the type of scenario discussed
  by Smith \& Owocki (2006).  Panel (a) shows the case for an initial
  stellar mass of 120 M$_{\odot}$ (appropriate for an LBV like AG
  Carinae), and Panel (b) shows an initial mass of 60 M$_{\odot}$
  (appropriate for P Cygni, perhaps).  The clumping factors of 4--6
  shown here are still fairly modest compared to some estimates of
  $>$10 for O-star winds.}
\end{center}
\end{figure}

% FIGURE 2 ----------
\begin{figure}[!ht]\begin{center}
\epsfig{file=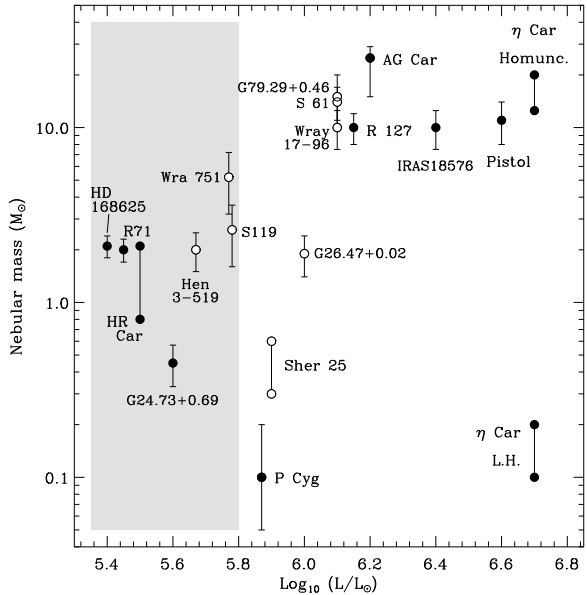,width=3.3in}
%\epsscale{0.99}
%\plotone{lbvneb.eps}
\caption{Masses of ejecta nebulae around LBVs (filled dots) and LBV
  candidates (unfilled) as a function of the central star's bolometric
  luminosity.  Luminosities are taken from Smith, Vink, \& de Koter
  (2004), while sources for the masses are given in Smith \& Owocki
  (2006).}
\end{center}
\end{figure}

\section{Alternative Scenarios}

The scenario where LBV eruptions dominate the mass loss of the most
massive stars (Fig.\ 1), would represent a dramatic change in our
understanding of mass loss in stellar evolution. The need for
recognizing the role of LBV eruptions in mass loss is partly motivated
by recent studies of the mass-loss rates of O stars, where wind
clumping suggests drastic reductions in $\dot{M}$ on the MS.  To be
fair, the amount of reduction in $\dot{M}$ is not yet settled; some
indications favor reduction of more than an order of magnitude, while
other estimates indicate factors of only a few (see the talk by J.\
Puls).  While this is debated, it is worth remembering that even if
the $\dot{M}$ reduction is only a factor of 3, {\it LBV eruptions may
still dominate the total mass lost during the lifetime of a very
massive star}.  The plots in Figure 2 adopt fairly modest mass-loss
rate reduction factors.

% FIGURE 3 ----------
\begin{figure}[!ht]\begin{center}
\epsfig{file=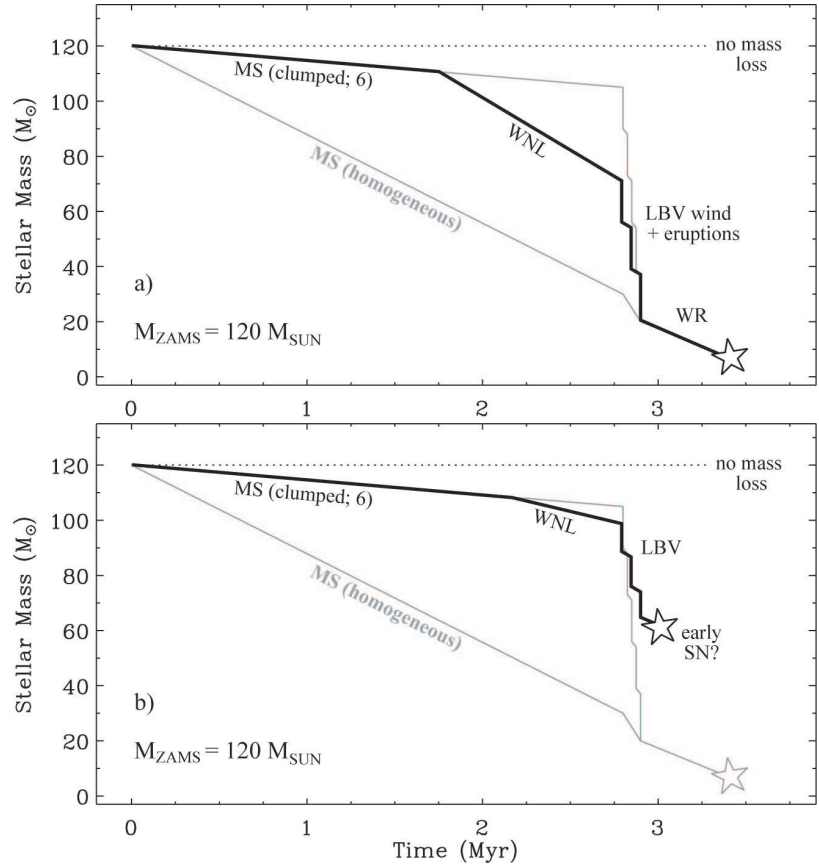,width=4.2in}
%\epsscale{0.99}
%\plotone{lbvneb.eps}
\caption{Same as Figure 1, but showing more complicated (and probably
  more realistic) hypothetical alternatives in between the two
  extremes of Figure 1$a$ (only one case of M$_{\rm ZAMS}$=120
  M$_{\odot}$ is shown here).  Panel (a) allows for substantial mass
  loss via a wind during an extended late-type WN phase that lasts for
  almost half the MS lifetime.  The total mass lost by the WNL wind is
  almost as much as through LBV eruptions, but one could easily adjust
  these depending on the duration of the WNL phase.  Panel (b) has a
  weaker WNL phase and allows for the possibility that the very most
  massive stars might explode before reaching the WR phase, thereby
  reducing the amount of mass that needs to be shed through LBV
  eruptions.}
\end{center}
\end{figure}

However, clumping in O-star winds is only part of the story.  The
other element is the observational reality that LBV eruptions like
$\eta$ Car's massive 19th century outburst do indeed occur, and we
have evidence that they occur more than once.  A star's mass budget
needs to allow for that.  However, if we require several 10's of solar
masses in LBV eruptions, plus enhanced mass loss during a WNL phase
(see below), we run into a {\it serious} problem --- homogeneous winds
simply do not allow enough room for additional mass loss through WNL
phases and LBV eruptions!  Thus, the mass-loss rates implied by the
assumption of homogeneous winds are not viable. I would then suggest
that the existence of WNL and LBV mass loss is an independent argument
that O star winds {\it must} be clumped, reducing $\dot{M}$ by at
least a factor of 2--3.

Of course, it is likely that neither extreme in Figure 1 is exactly
right.  The truth may lie somewhere in between, so let's consider two
likely alternatives.

\subsection{A Long WNL Phase?}

One alternative is that a very massive star spends a good fraction of
its H-burning MS lifetime as a late-type WN star (WNL; see Crowther et
al.\ 1995).  Even if their winds are clumped, WNL stars have higher
mass-loss rates than their O star counterparts.  Thus, if massive
stars can spend something like a third or half of their MS lifetime as
a WNL star, they can take a substantial chunk out of the star's total
mass.  This could temper the burden placed upon LBVs.  This scenario
is sketched in Figure 3$a$.  To me, something like Figure 3$a$ seems
to be a ``best bet'', but there are a few caveats to keep in mind.

First, Figure 3$a$ with its rather long WNL phase should only apply to
the most freakishly massive stars, with initial masses above roughly
90--100 M$_{\odot}$.  The justification for this comment is that
spectral type O3 and even O2 stars still exist in clusters within star
forming regions that are 2.5--3 Myr old (like Tr16 in the Carina
Nebula).  O3 stars probably have initial masses around 80--100
M$_{\odot}$ or so, and MS lifetimes around 3 Myr.  Therefore, these
stars cannot spend a substantial fraction of their H-burning lifetime
as a WNL star, because they evidently live for about 3 Myr without yet
reaching the WNL phase (ask P.\ Conti for an alternative hypothesis).
Only for the most massive stars might a relatively long WNL phase be
possible.  This makes me wonder if we have yet another a dichotomy in
stellar evolution, with different evolutionary sequences above and
below 100--120 M$_{\odot}$ -- much like the dichotomy above and below
45--50 M$_{\odot}$.  One could certainly make the case that the most
luminous evolved stars that are sometimes called LBVs or LBV
candidates -- stars like $\eta$~Car, the Pistol star, HD5980, and
possibly LBV~1806-20 -- have followed a different path than the
``normal'' LBVs like AG~Car and R127. Below this threshhold, hot
supergiants like Ofpe/WN9 and B[e] stars might fill similar roles.

Second, while I admit that WNL stars make a substantial contribution
to $\dot{M}$, their influence must be limited.  They cannot provide
the majority of mass lost by these stars, so the LBV eruption mass
loss must still dominate.  The reasoning behind this comment has to do
with the available mass budget of $\eta$~Carinae; namely, that $\eta$
Car is probably a post-WNL star, {\it but it still has retained most
of its original mass}.

% FIGURE 4 ----------
\begin{figure}[!ht]\begin{center}
\epsfig{file=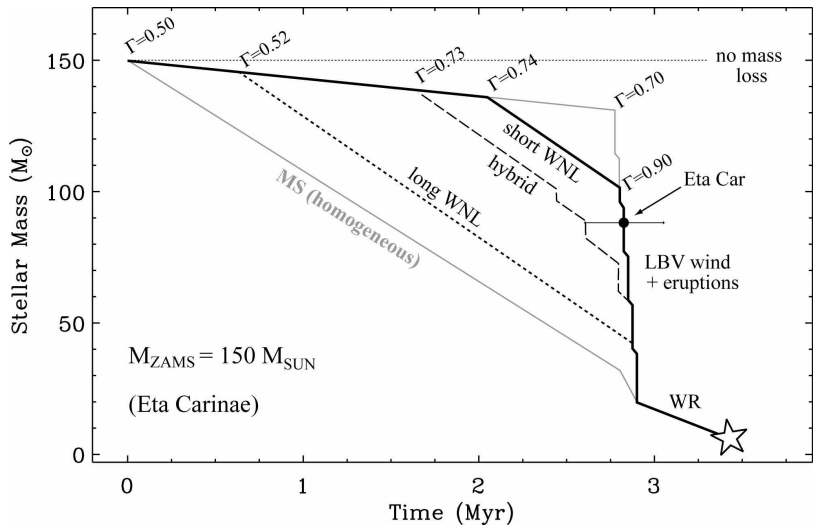,width=4.2in}
%\epsscale{0.99}
%\plotone{lbvneb.eps}
\caption{An artist's conception of the mass-loss history for a star
  with an initial mass at the upper limit of 150 M$_{\odot}$, perhaps
  appropriate for $\eta$ Carinae.  Here I show relatively short (small
  contribution) and a relatively long (dominant contribution) WNL
  phases, a ``hybrid'' WNL/LBV scenario, and the simpler extremes in
  gray.  Representative Eddington factors along the way are indicated.
  The dot shows the likely currently-observed locus of $\eta$ Carinae
  (note that I am being quite generous here with the correction for
  $\eta$ Car's companion star).  Considering that $\eta$ Car has
  already suffered 2--3 major LBV eruptions, which scenarios are
  consistent with its present mass?}
\end{center}
\end{figure}

Let's remember that $\eta$ Car is the most luminous and most evolved
member of a rich region containing over 65 O-type stars, as well as 3
WNL stars (see Smith 2006).  The current LBV phase of $\eta$ Car is
not only a post-MS phase, but probably also a post-WNL phase, since
its ejecta are more nitrogen rich than the WNL stars in Carina.  It is
also reasonable to assume that $\eta$ Car has advanced further in its
evolution sooner than the WNL stars of the same age in this region
simply because it is more luminous and started with a higher initial
mass.  Now, $\eta$ Car is seen today surviving as a very massive star
of around 100 M$_{\odot}$, and we measure a total of something like
20-35 M$_{\odot}$ in its circumstellar material ejected in only the
last few thousand years (the Homunculus, plus more extended outer
material; see Smith et al.\ 2003, 2005).  That means $\eta$ Car began
its LBV phase -- and ended its MS and WNL phases -- with $\sim$120
M$_{\odot}$ still bound to the star!  If there really is an upper
limit of about 150 M$_{\odot}$ to the mass of stars (Figer 2005;
Kroupa 2005), then {\it this rules out the possibility that winds
during the MS or WNL phases could dominate the mass-lost by the star
in its lifetime}.

This argument is made graphically in Figure 4, with options of
``long'' and ``short'' WNL phases.  Keeping three facts in mind ---
{\bf 1.} that we see more than 20 M$_{\odot}$ of nebular material from
recent LBV eruptions around $\eta$ Car, {\bf 2.} that $\eta$ Car has a
present day mass around 100 M$_{\odot}$ if it is not violating the
classical Eddington limit (I am being generous with the companion
star's mass in Figure 4), and {\bf 3.} that there is a likely upper
mass limit for stars of around 150 M$_{\odot}$ --- where could you
place $\eta$ Carinae on each track in Figure 4?  What does that
signify for the relative importance of the WNL phase?

Another not-too-ridiculous possibility may be the following: {\it What
if the WNL and LBV phases overlap, so that WNL stars are quiescent
LBVs for part of their existence}?  This scenario is shown by the
``hybrid'' track in Figure 4.  We already know that some LBVs (like AG
Car) are classified as Ofpe/WN9 stars in quiescence and make the
transition between the two states (Stahl 1986).  Perhaps the more
luminous WNL stars are also dormant LBVs for a spell.  In any case,
WNL stars do exist and their line-driven winds must play some role in
stellar evolution; the relative contribution of WNL vs.\ LBV eruptions
is a matter of degree, depending on the lifetime of the WNL phase.

Perhaps the most interesting consequence of the WNL mass loss is that
{\bf the WNL phase may facilitate the onset of the LBV instability}.
By quickly reducing the star's mass and thereby raising the star's L/M
ratio, the WNL wind may bring the star to the critical point where it
is dangerously close enough to the Eddington limit (say
$\Gamma\simeq$0.9) such that the LBV instability kicks in and takes
over the star's mass loss.  Representative Eddington factors are
indicated along the various tracks in Figure 4, and make the point
vividly.  This raises interesting questions about what happens at very
low metallicity, since the line-driven WNL wind should be weaker.

\subsection{An Early Death at the End of the LBV Phase?}

The main motivation for such huge amounts of mass loss in
continuum-driven LBV eruptions is the assumption that even the most
massive stars eventually need to reach the WR phase, requiring that
their mass be whittled down to about 20 M$_{\odot}$ before then (see
Smith \& Owocki 2006).  If we can relax this constraint so that the
most massive stars above 100 M$_{\odot}$ perhaps {\it do not} make it
to the WR phase, then we can alleviate the burden of removing so much
mass through LBV explosions.  In other words, {\it the most massive
stars might undergo core collapse at the end of the LBV phase, instead
of entering the WR phase} (Figure 3$b$).  If this scenario were true,
of course, it would mean that $\eta$~Carinae and stars like it in
other galaxies may explode as hypernovae at any moment.

We should be mindful that this alternative would require an {\it even
more} radical paradigm shift in our understanding of stellar evolution
than Figure 1$a$.  Namely, Figure 3$b$ would require that not only are
LBVs in core He burning, but that LBVs reach more advanced stages.
Yet, there are some reasons why an early explosion like in Figure 3$b$
might be attractive:

\begin{itemize}

\item As noted earlier in \S 2, observations of SNe (especially type
  IIn) and GRBs reveal that some have dense, massive circumstellar
  shells.  Where could these shells have come from if the WR phase has
  a sustained fast wind for a few 10$^5$ years?  The answer may be
  that these shells did in fact originate in LBV outbursts that
  occured within about 1000 years of the final death of the star.
  That would be astonishing and very important if true.

%\item Is there any example of a bona-fide WR star that is surrounded
%  by an extremely massive (like $\sim$50 M$_{\odot}$) group of nested
%  shells left over from a previous LBV phase?  If the most massive
%  stars explode at the end of the LBV phase, then we wouldn't
%  necessarily expect such massive shells.
 
\item In his talk at this meeting, Jorick Vink drew a similar
  conclusion about supernovae occurring at the end of the LBV phase,
  based on the radio evolution of objects like SN2001ig (e.g., Ryder
  et al.\ 2004).

\item A.\ Gal-Yam et al.\ (2007; in prep.) have identified a likely
  LBV as the progenitor for the Type IIn SN2005gl, and there may be
  others.

\end{itemize}

\section{Eruptive Mass Loss at Low Metallicity}

Unlike line-driven winds, the driving mechanism for giant LBV
outbursts is probably insensitive to metallicity (see Smith \& Owocki
2006).  There is good empirical evidence for this: Above
$\sim$10$^{5.8}$ L$_{\odot}$, no RSGs are seen because their redward
evolution is halted by heavy mass loss in the LBV phase (see many
contributions in the pre-fire Lunteren meeting).  This upper limit to
RSGs seems to hold even in low Z environments like the SMC (Humphreys
\& Davidson 1979), implying that the LBV instability is indeed
metallicity-independent.

The first stars, which were metal free, are thought to have been
predominantly massive (e.g., Bromm \& Larson 2004).  With no metals,
these stars should not have been able to launch line-driven winds, and
thus, they are expected to have suffered no mass loss during their
lifetimes.  The lack of mass loss profoundly affects the star's
evolution, the type of supernova it eventually produces (Heger et al.\
2003), and its yield of chemical elements.

This view rests upon the assumption that mass loss in massive stars at
the present time is dominated by line-driven winds, but this
assumption may be problematic because of the role of LBV outbursts and
their metallicity independence.  Furthermore, LBV mass loss in the
first stars might enable a WR phase to occur, wherein the star could
shed further mass through a line-driven wind because of
self-enrichment (Vink \& de Koter 2005; Eldridge \& Vink 2006). If
mass loss of massive stars at the present epoch is dominated by a
mechanism that is insensitive to metallicity, then we must question
the prevalent notion that the first stars suffered no mass loss before
their final SN event.  If these outbursts can occur at low
metallicity, it would profoundly alter our understanding of the
evolution of the first stars and their role in early galaxies.

\acknowledgments

I thank Stan Owocki, Paul Crowther, and Peter Conti for many relevant
discussions.  I was supported by NASA through grant HF-01166.01A from
STScI.

%%% THE BIBLIOGRAPHY
% REFERENCES


\begin{thebibliography}{}

\bibitem[]{} Appenzeller, I.\ 1986, IAU Symp.\ 116, 139

\bibitem[]{} Bohannan, B.\ 1997, in ASP Conf.\ Ser.\ 120, 3

\bibitem[]{} Bouret, J.C., Lanz, T., \& Hillier, D.J.\ 2005, A\&A,
438, 301

\bibitem[]{} Bromm, V., \& Larson, R.B.\ 2004, ARAA, 42, 79

\bibitem[]{} Crowther, P.A., et al.\ 2002, ApJ, 579, 774

\bibitem[]{} Crowther, P.A., et al.\ 1995, A\&A, 293, 427

\bibitem[]{} Davidson, K.\ 1987, in Instab.\ in Lum.\ Early-type Stars
  (Dordrecht: Reidel), 127

\bibitem[]{} Eldridge, J.J., \& Vink, J.S.\ 2006, A\&A, 452, 295

\bibitem[]{} Evans, C.J., et al.\ 2004, ApJ, 610, 1021

\bibitem[]{} Figer, D.F.\ 2005, Nature, 434, 192

\bibitem[]{} Fullerton, A.W., Massa, D.L., \& Prinja, R.K.\ 2006, ApJ,
637, 1025

\bibitem[]{} Heger, A., et al.\ 2003, ApJ, 591, 288

\bibitem[]{} Hillier, D.J., Lanz, T., Heap, S.R., et al.\ 2003, ApJ,
588, 1039

\bibitem[]{} Humphreys, R.M., \& Davidson, K.\ 1979, ApJ, 232, 409

\bibitem[]{} Humphreys, R.M., \& Davidson, K.\ 1994, PASP, 106, 1025

\bibitem[]{} Humphreys, R.M., Davidson, K., \& Smith, N.\ 1999, PASP,
111, 1124

\bibitem[]{} Kroupa, P.\ 2005, Nature, 434, 148

\bibitem[]{} Lamers, H.J.G.L.M.\ 1987, in Instab.\ in Lum.\ Early-type
  Stars (Dordrecht: Reidel), 99

\bibitem[]{} Lamers, H.J.G.L.M., \& Fitzpatrick, E.\ 1988, ApJ, 324, 279

\bibitem[]{} Massa, D., et al.\ 2003, ApJ, 586, 996

\bibitem[]{} Meaburn, J.\ 2001, in ASP Conf.\ Ser.\ 233, 253

\bibitem[]{} Morse, J.A., et al.\ 2001, ApJ, 548, L207

\bibitem[]{} Owocki, S.P., Gayley, K.G., \& Shaviv, N.J.\ 2004, ApJ,
616, 525

\bibitem[]{} Puls, J., et al.\ 2006, A\&A, 454, 625

\bibitem[]{} Ryder, S.D., et al.\ 2004, MNRAS, 349, 1093

\bibitem[]{} Smith, N.\ 2006, MNRAS, 367, 763

\bibitem[]{} Smith, N., \& Hartigan, P.\ 2006, ApJ, 638, 1045

\bibitem[]{} Smith, N., \& Owocki, S.P.\ 2006, ApJ, 645, L45

\bibitem[]{} Smith, N., et al.\ 2003, AJ, 125, 1458

\bibitem[]{} Smith, N., Morse, J.A., \& Bally, J.\ 2005, AJ, 130, 1778

\bibitem[]{} Smith, N., Vink, J., \& de Koter, A.\ 2004, ApJ, 615, 475

\bibitem[]{} Stahl, O.\ 1986, A\&A, 164, 321

\bibitem[]{} Vink, J.S., \& de Koter, A.\ 2005, A\&A, 442, 587

\bibitem[]{} Walborn, N.R.\ 1976, ApJ, 204, L17

\end{thebibliography}
\end{document}